\documentclass[twocolumn,secnumarabic,amssymb, amsmath, nobibnotes, aps, prd]{revtex4-1}

\usepackage[dvipdfmx]{graphicx}  
\usepackage{dcolumn}   
\usepackage{bm}        
\usepackage{amssymb}   
\usepackage{here}
\usepackage{amsmath}
\usepackage{comment}
\usepackage{color}

\usepackage{braket}
\usepackage{mathrsfs}
\usepackage{amsmath}
\usepackage{ulem}
\usepackage{fancybox}
\usepackage{xspace}
\usepackage{amsfonts}
\usepackage{multirow}
\usepackage{amsthm}
\usepackage{setspace}
\usepackage[subrefformat=parens]{subcaption}

\begin{document}

\title{${\mathbb Z}_2$ topological invariant for magnon spin Hall systems}%

\author{Hiroki Kondo, Yutaka Akagi, and Hosho Katsura}
\affiliation{Department of Physics, Graduate School of Science, The University of Tokyo, Hongo, Tokyo 113-0033, Japan}
\email[]{kondo-hiroki290@g.ecc.u-tokyo.ac.jp}
%\date{today}%

\begin{abstract}
We propose a definition of a ${\mathbb Z}_2$ topological invariant for magnon spin Hall systems which are the bosonic analog of two-dimensional topological insulators in class AII. 
The existence of ``Kramers pairs" in these systems is guaranteed by pseudo-time-reversal symmetry which is the same as time-reversal symmetry up to some unitary transformation. The ${\mathbb Z}_2$ index of each Kramers pair of bands is expressed in terms of the bosonic counterparts of the Berry connection and curvature. 
We construct explicit examples of magnon spin Hall systems and demonstrate that our ${\mathbb Z}_2$ index precisely characterizes the presence or absence of helical edge states.
The proposed ${\mathbb Z}_2$ index and the formalism developed can be applied not only to magnonic systems but also to other non-interacting bosonic systems. 
\end{abstract}
\maketitle

%Introduction
{\it Introduction.}
The classification and characterization of different phases of matter based on the topology of band structures has recently attracted considerable attention~\cite{Schnyder08, Kitaev09, Ryu10}. 
In general, different phases are distinguished by their topological invariants. 
The most famous example of such a topological invariant is the first Chern number~\cite{Thouless82, Kohmoto85}, 
which is in one-to-one correspondence with the number of chiral edge states~\cite{Hatsugai93a, Hatsugai93b} in quantum Hall systems~\cite{Klitzing80}. 
Over the past decade, it has been recognized that this bulk-edge correspondence is not limited to systems with broken time-reversal symmetry. 
Time-reversal symmetry and other discrete symmetries inherent in crystals lead to a variety of new topological invariants and a more refined classification of phases, which are protected as long as such symmetries are preserved~\cite{Chiu13, Morimoto13, Fang12, Alexandradinata14, Shiozaki15a, Liu14, Fang15, Shiozaki15b, Shiozaki16, Wang16, Shiozaki17, Kruthoff17, Po17, Bradlyn17, Watanabe17}.

The prime examples of topological phases protected by time-reversal symmetry are two- and three-dimensional topological insulators in class AII~\cite{Kane05a,Kane05b, Hasan10, Qi11}. 
These topological insulators possess a helical edge state which carries electrons with opposite spins propagating in opposite directions. 
In two dimensions, this results in the spin Hall effect. 
The topological invariant that characterizes the presence or absence of a helical edge state is called the $\mathbb{Z}_2$ index. A model of a topological insulator with a nontrivial $\mathbb{Z}_2$ index was theoretically proposed by combining two copies of the Haldane model~\cite{Haldane88} so that the total system restores time-reversal symmetry~\cite{Kane05a, Kane05b}.

The successful studies on the topological phases of electrons have been extended to bosonic systems with fascinating phenomena such as the photon~\cite{Onoda04, Hosten08, Raghu08, Haldane08, Wang09a, Ben-Abdallah16}, phonon~\cite{Strohm05, Sheng06, Inyushkin07, Kagan08, Wang09b, Zhang10, Qin12, Mori14, Sugii17}, magnon~\cite{Fujimoto09, Katsura10, Matsumoto11, Matsumoto14, Shindou13a, Shindou13b, Kim16, Onose10, Ideue12,Chisnell15, Hirschberger15, Han_Lee17, Murakami_Okamoto17}, and triplon Hall effects~\cite{Rumhanyi15}. 
Of particular interest are magnons, which are the quasiparticles of spin waves. Magnons could be observed in real time and space in experiments~\cite{Demokritov01} and have potential applications in spintronics, as they have a long coherence and carry angular momenta. 
%~\cite{Buttner00}.
In fact, the intrinsic topological (thermal) Hall effect of magnons was confirmed by experiments~\cite{Onose10, Ideue12,Chisnell15, Hirschberger15} immediately after the theoretical predictions~\cite{Fujimoto09, Katsura10, Matsumoto11, Matsumoto14, Shindou13a, Shindou13b, Kim16}.
This leads us to expect that theoretical proposals in this field will find an experimental realization in a short period of time. 
To date, a variety of novel phenomena and states of matter related to magnons have been proposed. Examples include the magnon spin Nernst effect~\cite{Cheng16, Zyuzin16, Nakata17, Mook18}, and  Dirac~\cite{Fransson16,Owerre17} and Weyl magnons~\cite{Li16, Mook16, Su17}.

Spin systems exhibiting the magnon spin Nernst effect without the thermal Hall effect~\cite{Zyuzin16, Nakata17, Mook18} can be regarded as a magnon analog of ${\mathbb Z}_2$ topological insulators, where a helical magnon edge state is expected to exist. 
In this Rapid Communication, we refer to such systems as magnon spin Hall (MSH) systems. 
The topological invariant that characterizes the presence or absence of edge states in MSH systems has not been fully identified. This is because time-reversal symmetry does not necessarily imply a Kramers degeneracy for bosonic systems. The only exceptions are systems in which the $z$ component of the spin $S_z$ (along the N\'eel vector) is conserved and the spin Chern number is well defined~\cite{Nakata17}.

In this Rapid Communication, we define a ${\mathbb Z}_2$ index that characterizes the MSH systems by constructing Kramers pairs in bosonic systems.
To demonstrate the validity of the $\mathbb{Z}_2$ index, we study ``ferromagnetic" kagome and antiferromagnetic honeycomb bilayer systems. 
The latter system can be thought of as a magnetic analog of the Kane-Mele model~\cite{Kane05a, Kane05b}. This is an example of MSH systems without the conservation of $S_{z}$. 
In both cases, we show numerically that the ${\mathbb Z}_2$ index characterizes the presence or absence of  edge states and remains robust against small changes in the parameters.

%Set-up
{\it Definition of $\mathbb{Z}_2$ topological invariant.}
We start from a system of noninteracting bosons in two dimensions. Assuming translational invariance, a generic quadratic Hamiltonian describing the system is given by
\begin{align}
\mathcal{H}=\frac{1}{2}\sum_{\bm{k}}[\bm{\beta}^{\dagger}(\bm{k})\bm{\beta}(-\bm{k})]H(\bm{k})
\left[
\begin{array}{cc}
\bm{\beta}(\bm{k})  \\
\bm{\beta}^{\dagger}(-\bm{k})  \\
\end{array} 
\right].
\label{eq:gene_Ham}
\end{align}
Here, $\bm{\beta}^{\dagger}(\bm{k})=[\beta_{1}^{\dagger}(\bm{k}),\dots,\beta_{\mathscr{N}}^{\dagger}(\bm{k})]$ denotes boson creation operators with momentum $\bm{k} = (k_x, k_y)$.
The subscript $\mathscr{N}$ is the number of internal degrees of freedom in a unit cell, which we assume to be even. 
The $2\mathscr{N}\times2\mathscr{N}$ Hermitian matrix $H(\bm{k})$ is a bosonic Bogoliubov--de Gennes (BdG) Hamiltonian.

In order to construct the bosonic analog of the class AII topological insulator, we introduce a pseudo-time-reversal operator $\Theta'=PK$, where $P$ is a ${\bm k}$-independent paraunitary matrix and $K$ is the complex conjugation. 
The operator $\Theta'$ and $P$ satisfy the following relations,
\begin{align}
&\Theta'^{2}=-1 \label{eq:Theta},\\
&P^{\dagger}\Sigma_{z}P=\Sigma_{z} \label{eq:P}.
\end{align}
Here, $\Sigma_{z}$ is defined as a tensor product $\Sigma_{z}:=\sigma_z \otimes  1_\mathscr{N}$, where $\sigma_a$ ($a=x,y,z$) is the $a$ component of the Pauli matrix acting on the particle-hole space and $1_\mathscr{N}$ is the $\mathscr{N} \times \mathscr{N}$ identity matrix. 
We require that a bosonic BdG Hamiltonian with pseudo-time-reversal symmetry meets
\begin{align}
\Sigma_{z}H(-\bm{k})\Theta'-\Theta'\Sigma_{z}H(\bm{k})=0. 
\label{eq:Commutation}
\end{align}
Note that the operator $\Theta'$ satisfies Eq.~(\ref{eq:Theta}) as in fermionic systems, while
the conventional time-reversal operator~\cite{comment_TRS} squares to $+1$ for bosonic systems. 
Explicit expressions for $\Theta'$ and $H({\bm k})$ will be given later in Eqs. (\ref{eq:pseudoTR}) and (\ref{eq:H}).  
%We will give a concrete expression of $\Theta'$ later.
The operator $\Theta'$ ensures the existence of ``Kramers pairs" of bosons~\cite{Wu15, He16, Ochiai15} owing to Eq.~(\ref{eq:Theta}) and a nontrivial inner product for bosonic wavefunctions defined as
\begin{align}
\langle \! \langle \bm{\phi},\bm{\psi} \rangle \! \rangle=\bm{\phi}^{\dagger}\Sigma_{z}\bm{\psi},
\end{align}
where $\bm{\phi}$ and $\bm{\psi}$ are $2\mathscr{N}$-dimensional complex vectors and $\bm{\phi}^{\dagger}$ is the adjoint of $\bm{\phi}$.
See Supplemental Material for the proof of existence of  Kramers pairs.

Next, we introduce a definition of the ${\mathbb Z}_2$ topological 
invariant for bosonic systems with pseudo-time-reversal symmetry. 
For fermionic systems, there are various definitions of the ${\mathbb Z}_2$ invariant~\cite{Fu06, Fukui07, Moore07, Fukui08, Qi08, Roy09, Wang10, Loring10, Fulga12, Sbierski14, Loring15, Katsura16, Akagi17, Katsura18}. 
Here, we follow the formalism developed by Fu and Kane~\cite{Fu06}.

Let ${\bm \Psi}_{n,1,+}({\bm k})$ ($n=1,\dots,{\mathscr N}/2$) be an eigenvector of $\Sigma_z H({\bm k})$ with an eigenvalue $E_n({\bm k}) \ge 0$, i.e., a particle wave function. The normalization is fixed by the condition $\langle \! \langle {\bm \Psi}_{n,1,+} ({\bm k}), {\bm \Psi}_{n,1,+} ({\bm k}) \rangle \! \rangle = 1$. It follows from Eq. (\ref{eq:Commutation}) that ${\bm \Psi}_{n,2,+}({\bm k}):=-\Theta' {\bm \Psi}_{n,1,+} (-{\bm k})$ is an eigenvector of $\Sigma_z H({\bm k})$ with an eigenvalue $E_n(-{\bm k})$. The $n$th Kramers pair of bands is formed by ${\bm \Psi}_{n,l,+} ({\bm k})$ ($l=1,2$). The Kramers degeneracy at a time-reversal invariant momentum follows from the property of $\Theta'$, i.e., Eq. (\ref{eq:Theta}). The particle-hole conjugates can be obtained as ${\bm \Psi}_{n,l,-}({\bm k}) = \Sigma_x K {\bm \Psi}_{n,l,+}(-{\bm k})$, where $\Sigma_x:=\sigma_x \otimes 1_{\mathscr N}$. They are the eigenvectors of $\Sigma_z H({\bm k})$ with an  eigenvalue $-E_n((-1)^l {\bm k})$ and satisfy $\langle \! \langle {\bm \Psi}_{n,l,-} ({\bm k}), {\bm \Psi}_{n,l,-} ({\bm k}) \rangle \! \rangle = -1$.

The Berry connection and curvature for the $n$th Kramers pair of particle (hole) bands is defined as
\begin{align}
&\bm{A}_{n, \sigma}(\bm{k})=\sum_{l=1,2}\bm{A}_{n,l,\sigma}(\bm{k}),
\\
&\Omega_{n,\sigma}(\bm{k})=\sum_{l=1,2}\Omega_{n,l,\sigma}(\bm{k}), 
\end{align}
where 
\begin{align}
&\bm{A}_{n,l,\sigma}(\bm{k})
={\rm i}\, \sigma \left\langle \! \left\langle \bm{\Psi}_{n,l,\sigma}(\bm{k}),\nabla_{\bm{k}}\bm{\Psi}_{n,l,\sigma}(\bm{k})\right\rangle \! \right\rangle,
\\
&\Omega_{n,l,\sigma}(\bm{k})=\bigl[ \nabla_{\bm{k}} \times \bm{A}_{n,l,\sigma}(\bm{k}) \bigr]_{z}.
\end{align} 
Here, $[\: \cdot \:]_z$ represents the $z$ component of the three-dimensional vector in the brackets. 
The Berry connections of the particle bands and those of the hole bands are related to each other via ${\bm A}_{n,1,+}({\bm k})={\bm A}_{n,2,-}({\bm k})$ and ${\bm A}_{n,2,+}({\bm k})={\bm A}_{n,1,-}({\bm k})$, yielding ${\bm A}_{n,\sigma}({\bm k})={\bm A}_{n,-\sigma}({\bm k})$ and $\Omega_{n,\sigma}({\bm k})=\Omega_{n,-\sigma}({\bm k})$. 
%The Berry connection and curvature of the hole Kramers pair and those of the particle Kramers pair are related to each other via ${\bm A}_{n,\sigma}({\bm k})={\bm A}_{n,-\sigma}({\bm k})$ and $\Omega_{n,\sigma}({\bm k})=\Omega_{n,-\sigma}({\bm k})$, which follow from ${\bm A}_{n,1,+}({\bm k})={\bm A}_{n,2,-}({\bm k})$ and ${\bm A}_{n,2,+}({\bm k})={\bm A}_{n,1,-}({\bm k})$.

\begin{figure}[H]
\centering
  \includegraphics[width=0.97\columnwidth]{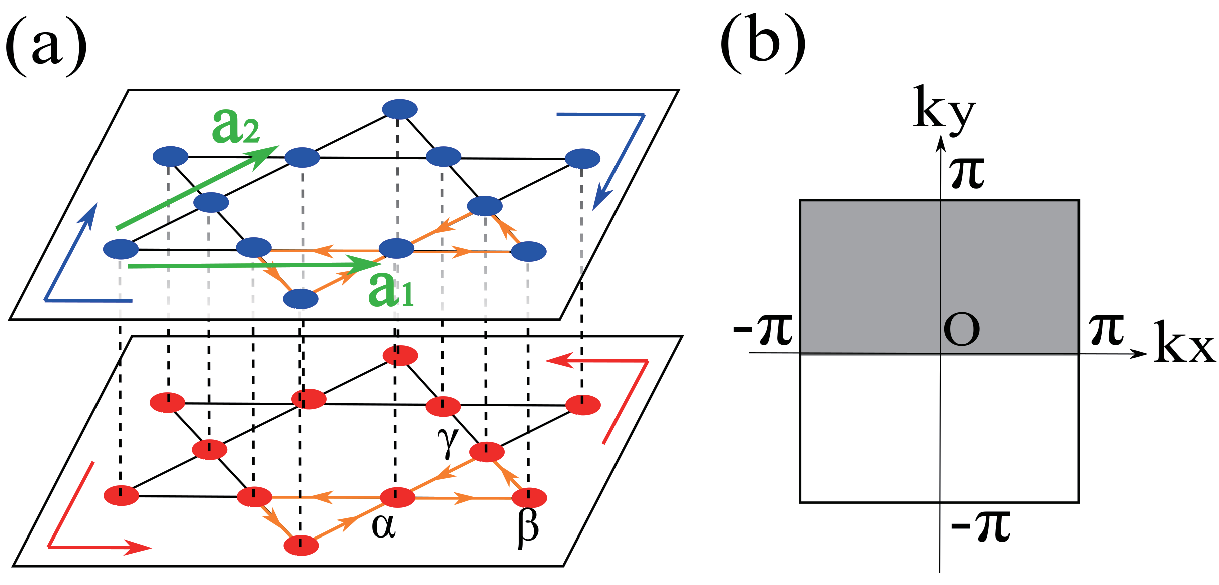}
\caption{(Color online) (a) The bilayer kagome system. 
The red and blue dots indicate spins pointing in the $+z$ and $-z$ directions, respectively.
The vectors $\bm{a}_{1}$ and $\bm{a}_{2}$ are the primitive lattice vectors. 
The three sublattices of each kagome layer are indicated by $\alpha$, $\beta$, and $\gamma$.
The orange arrows indicate the sign convention for the DM vectors. 
Magnon edge states with opposite magnetic dipole moments propagate in opposite directions, as shown by the red and blue arrows.
(b) The Brillouin zone (BZ) and the effective Brillouin zone (EBZ) indicated by the shaded region. 
 Taking $\bm{a}_{1}=(1,0)$ and $\bm{a}_{2}=(0,1)$, we here deform the kagome lattice into an equivalent square lattice. 
}\label{fig:kagome}
\end{figure}

Using $\bm{A}_{n,\sigma}$ and $\Omega_{n,\sigma}$, the ${\mathbb Z}_2$ index of the $n$th Kramers pair of bands is defined as 
\begin{align}
D_{n,\sigma} \! := \frac{1}{2\pi} \! \left[ \oint_{\partial {\rm EBZ}} \!\!\!\!\!\!\!\!\!\!\!\! d\bm{k} \cdot \bm{A}_{n,\sigma}(\bm{k}) \! - \!\! \int_{\rm EBZ} \!\!\!\!\!\!\!\!\! d^2k \; \Omega_{n,\sigma}(\bm{k}) \right]\hspace{1mm}{\rm mod}\hspace{1mm}2, 
\label{eq:D}
\end{align}
where EBZ and ${\partial {\rm EBZ}}$ denote the effective Brillouin zone and its boundary, respectively. The EBZ related to the time-reversal-invariant band structures describes one half of the Brillouin zone [e.g., see Fig. \ref{fig:kagome} (b)]. 
The pseudo-time-reversal symmetry leads to the $\mathbb{Z}_2$ quantization of $D_{n,\sigma}$ for the same reason as in the fermionic case~\cite{Fu06}. Equation (\ref{eq:D}) is our main result. 
Since the relation $D_{n, \sigma} = D_{n, -\sigma}$ holds, we drop the subscript $\sigma=\pm$ in the following.

{\it Models.}
So far we have not specified the form of the Hamiltonian $H({\bm k})$. To demonstrate the validity of the $\mathbb{Z}_2$ index defined in Eq.~(\ref{eq:D}), we consider bilayer systems with collinear magnetic order. 
For such systems, the magnon creation operator $\bm{\beta}^{\dagger}(\bm{k})$ in Eq. (\ref{eq:gene_Ham}) can be generally written as
\begin{align}
{\bm{\beta}}^{\dagger}(\bm{k})=[\bm{b}_{\uparrow}^{\dagger}(\bm{k}),\bm{b}_{\downarrow}^{\dagger}(\bm{k})],
\end{align}
where
%The creation operators of magnons originating from up-spins 
$\bm{b}_{\uparrow}^{\dagger}(\bm{k})$ and 
%down-spins 
$\bm{b}_{\downarrow}^{\dagger}(\bm{k})$ are given by 
\begin{align}
&\bm{b}_{\uparrow}^{\dagger}(\bm{k})=
[b_{\alpha_{1},1}^{\dagger}(\bm{k}),\cdots,b_{\alpha_{N},1}^{\dagger}(\bm{k}),b_{\beta_{1},2}^{\dagger}(\bm{k}),\cdots,b_{\beta_{M},2}^{\dagger}(\bm{k})], \nonumber \\
&\bm{b}_{\downarrow}^{\dagger}(\bm{k})=
[b_{\alpha_{1},2}^{\dagger}(\bm{k}),\cdots,b_{\alpha_{N},2}^{\dagger}(\bm{k}),b_{\beta_{1},1}^{\dagger}(\bm{k}),\cdots,b_{\beta_{M},1}^{\dagger}(\bm{k})]. \nonumber
\end{align}
The operator $b_{\mu_{n},l}^{\dagger}(\bm{k})$ creates a magnon on the sublattice $\mu_{n}$ on the $l$th layer. 
Here, we assumed that the system has $N+M$ sublattices in each layer, namely, the spins on the sublattices $\alpha_{1},\dots,\alpha_{N}$ and $\beta_{1},\dots,\beta_{M}$ of the first (bottom) layer point in the $+z$ and $-z$ directions, respectively [the $N$ ($M$) spins point upward (downward) in a unit cell of the first layer]. 
The spins on the second (top) layer have the directions opposite to those on the first layer. 
Now we introduce the following pseudo-time-reversal operator,
\begin{align}
\Theta'=(\sigma_z \otimes {\rm i} \sigma_{y} \otimes 1_{N+M})K.
\label{eq:pseudoTR}
\end{align}
The part $\sigma_z$ acts on the particle-hole space, while ${\rm i}\sigma_y$ interchanges the top and bottom layers with an extra sign. 
With this $\Theta'$, the most general Hamiltonian satisfying Eq.~(\ref{eq:Commutation}) takes the form
\begin{align}
H(\bm{k})=
\left(
\begin{array}{cccc}
h_{1}(\bm{k}) &h_{2}(\bm{k}) &\Delta_{2}(\bm{k}) &\Delta_{1}(\bm{k}) \\
h_{2}^{\dagger}(\bm{k}) &h_{1}^{*}(-\bm{k}) &\Delta_{1}^{*}(-\bm{k})&-\Delta_{2}^{\dagger}(\bm{k}) \\
\Delta_{2}^{\dagger}(\bm{k}) &\Delta_{1}^{*}(-\bm{k}) &h_{1}^{*}(-\bm{k}) &h_{2}^{*}(-\bm{k}) \\
\Delta_{1}(\bm{k}) &-\Delta_{2}(\bm{k}) &h_{2}^{T}(-\bm{k}) &h_{1}(\bm{k}) \\
\end{array} 
\right),
\label{eq:H}
\end{align}
where $h_i(\bm{k})$ and $\Delta_i(\bm{k})$ for $i=1, 2$ are $(N+M) \times (N+M)$ matrices and satisfy $h_{1}^{\dagger}(\bm{k})=h_{1}(\bm{k}),\Delta_{1}^{\dagger}(\bm{k})=\Delta_{1}(\bm{k}),h_{2}^{T}(\bm{k})=-h_{2}(-\bm{k}),$ and $\Delta_{2}^{T}(\bm{k})=\Delta_{2}(-\bm{k})$. 
Intuitively, the symmetry of $H({\bm k})$ means that the system is invariant under the combination of time reversal and interchange of the two layers. 

%Figure 1: kagome bilayer
Our first explicit example of MSH systems is a ``ferromagnetic" bilayer kagome system without a net moment. This is the case of $N=3$ and $M=0$. 
Here, we assume that the spins on the same layer are aligned in the same direction, while those in different layers are aligned in opposite directions due to the interlayer antiferromagnetic coupling [see Fig.~\ref{fig:kagome} (a)]. 
To realize such a system, we consider the following Hamiltonian
\begin{align}
\mathcal{H}_{\rm{K}}=\mathcal{H}^{(1)}+\mathcal{H}^{(2)}+\mathcal{H}^{\rm (inter)}, 
 \label{eq:kagome}
\end{align}
where
\begin{align}
&\mathcal{H}^{(l)}=-J\sum_{\langle i,j\rangle} \bm{S}_{i}^{(l)}\cdot\bm{S}_{j}^{(l)}+D\sum_{\langle i,j\rangle} \xi_{ij}\left( \bm{S}_{i}^{(l)}\times\bm{S}_{j}^{(l)} \right)_z,
\nonumber \\
&\mathcal{H}^{\rm (inter)}=J'\sum_{i} \bm{S}_{i}^{(1)}\cdot\bm{S}_{i}^{(2)},
\nonumber
\end{align}
with $J, J' >0$. Here, $\bm{S}_{i}^{(l)}$ denotes the spin at site $i$ on the $l$th layer. 
The first and second terms of $\mathcal{H}^{(l)}$ represent the ferromagnetic Heisenberg and the Dzyaloshinskii-Moriya (DM) interactions between nearest-neighbor spins, respectively. 
%The sign convention of $\xi_{ij}$ is taken as $+1$ in the direction of the orange arrows in Fig.~\ref{fig:kagome} (a), while $\xi_{ij}=-1$ in the opposite direction of the orange arrows.
The sign convention of $\xi_{ij}$ is such that $\xi_{ij}=+1$ if the arrow of the link $\langle i, j \rangle$ points from $i$ to $j$ and $\xi_{ij}=-1$ if the arrow points from $j$ to $i$ [see the orange arrows in Fig.~\ref{fig:kagome} (a)]. 
Note that the Hamiltonian of each layer ${\cal H}^{(l)}$ describes a single-layer kagome system, which exhibits the magnon (thermal) Hall effect~\cite{Mook14, Seshadri18}. 
The remaining term $\mathcal{H}^{\rm (inter)}$ represents the interlayer antiferromagnetic Heisenberg interaction and the sum runs over the vertical spin pairs [see the dashed lines in Fig.~\ref{fig:kagome} (a)]. Assuming the aforementioned magnetic order and applying the Holstein-Primakoff transformation, the Hamiltonian (\ref{eq:kagome}) becomes the same as Eq.~(\ref{eq:H}) with 
\begin{align}
h_{1}(\bm{k})=
S\left(
\begin{array}{ccc}
4J+J' &-z p_{1}^{*} &-z^{*}p_{2}^{*} \\
-z^{*}p_{1} &4J+J' &-z p_{3}^{*}  \\
-z p_{2} &-z^{*}p_{3} &4J+J'  \\
\end{array} 
\right),
\label{eq:single_layer}
\end{align}
$h_{2}(\bm{k})=0$, $\Delta_{1}=J'S1_{3}$, and $\Delta_{2}(\bm{k})=0$. 
Here, $z=J+{\rm i}D$ and $p_i=1+\exp({\rm i}\,{\bm k}\cdot {\bm a}_i)$ with ${\bm a}_1=(1,0)$, ${\bm a}_2=(0,1)$, and ${\bm a}_3={\bm a}_2-{\bm a}_1$, as shown in Fig. \ref{fig:kagome} (a). 
In the limit of $J'\to 0$, the matrix $h_{1}(\bm{k})$ corresponds to the Hamiltonian of the single kagome layer.

%Figure 2: kagome edge states
\begin{figure}[H]
  \centering
  \includegraphics[width=0.97\columnwidth]{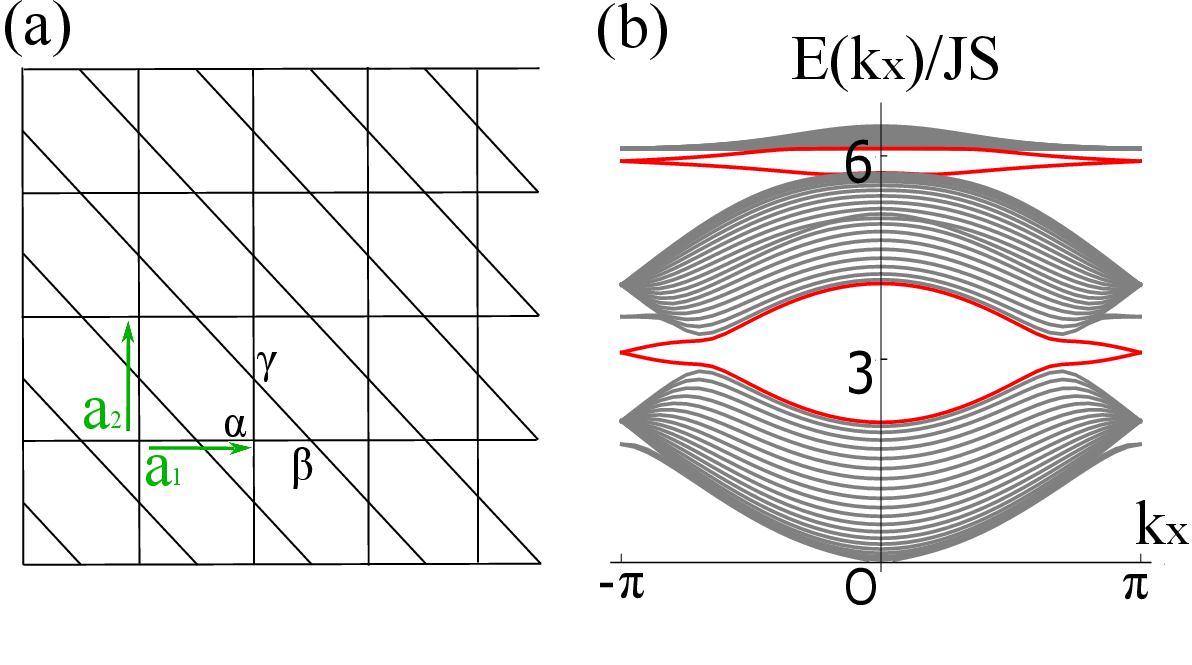}
\caption{(a) A kagome strip of width $M$ unit cells with periodic (open) boundary conditions in the horizontal (vertical) direction. 
(b) Magnon spectrum of the strip of the bilayer kagome lattice with $M=20$ width for $D=0.1 J$, $J'=0.1J$, and $J>0$. 
Topologically protected magnon helical edge states shown in red occur in each energy gap.}
    \label{fig:edge} 
\end{figure}

Figure~\ref{fig:edge}(b) shows the result of the magnon dispersion in the bilayer kagome system described by Eq.~(\ref{eq:kagome}) with cylindrical boundary conditions (Fig. \ref{fig:edge} (a)). 
We here plot only one of the Kramers pairs due to the double degeneracy of each band. 
 %in the language of the magnon Hamiltonian.
%The magnon excitation has two gapless modes and four gapped modes at the $\Gamma$ point.
The distinctive feature of the spectrum is the edge states (indicated by red color), which traverse the energy gaps. 
%~\cite{beard}
%The important point of the figure is that the edge states traverse between
Correspondingly, using Eq.~(\ref{eq:D}) and the numerical implementation described in Ref.~\cite{Fukui07}, we obtain that the ${\mathbb Z}_2$ indices are 1, 0, and 1 from the lowest band to the highest band, i.e., 
%associated with four bands are given by  1, 1, 1, and  1 from the lowest band to the highest band, respectively.
%$D_{1\sigma}=D_{3\sigma}=1$, and $D_{2\sigma}=0$ for $\sigma=\pm$.
$D_{1}=1$, $D_{2}=0$, and $D_{3}=1$. 
The indices remain the same by changing the parameters as long as the aforementioned magnetic order is stable. 
We also confirm that a topologically trivial phase with $D_{n}=0$ for each band is realized by adding a single-ion anisotropy term to the Hamiltonian Eq. (\ref{eq:kagome}) (see Supplemental Material for details).

\begin{table}[H]
\caption{
The relation between Chern numbers and $\mathbb{Z}_2$ indices. 
Here, $C_{n,1}$ ($C_{n,2}$) denotes the Chern number of the band labeled by $n,1$ $(n,2)$, while $D_{n}$ is the $\mathbb{Z}_2$ index of the $n$th Kramers pair of bands for the bilayer kagome system~\cite{Chern_hole}. 
Each Kramers pair with ${\mathbb Z}_2$ index $1$ consists of two bands with Chern numbers $+1$ and $-1$.}
\begin{center}
{\tabcolsep=5mm
  \begin{tabular}{lccc}
    \hline \hline
    $n$ & $C_{n,1}$ & $C_{n,2}$ & $D_{n}$ \\ \hline
    $1$ (top) & $+1$ & $-1$ & 1 \\ \hline
    $2$ (middle) & 0 & 0 & 0 \\ \hline
    $3$ (bottom) & $-1$ & $+1$ & 1 \\
    \hline \hline
  \end{tabular}
}
\end{center}
\label{table:chern}
\end{table}
As is clear from Table~\ref{table:chern}, the nontrivial values of the ${\mathbb Z}_2$ indices come from the pair of Chern numbers~\cite{comment_Chern} $+1$ and $-1$. 
%of the single kagome layers.
In fact, due to the specific form of the Hamiltonian, we can think of the ${\mathbb Z}_2$ index as the spin Chern number $D_{n}=\frac{1}{2}(C_{n,1}-C_{n,2})$ (mod $2$), as in electronic systems with conservation of $S_z$. 
Because of the pseudo-time-reversal symmetry, the total Chern number of each Kramers pair vanishes, i.e., $C_n=C_{n,1}+C_{n,2}=0$. 
We note in passing that the Berry connection and curvature of a system consisting of two antiferromagnetically coupled ferromagnetic layers perfectly coincide with those of the two independent single-layer systems without interlayer coupling (see Supplemental Material for details).

%Figure 3: bilayer honeycomb
\begin{figure}[H]
  \centering
  \includegraphics[width=0.97\columnwidth]{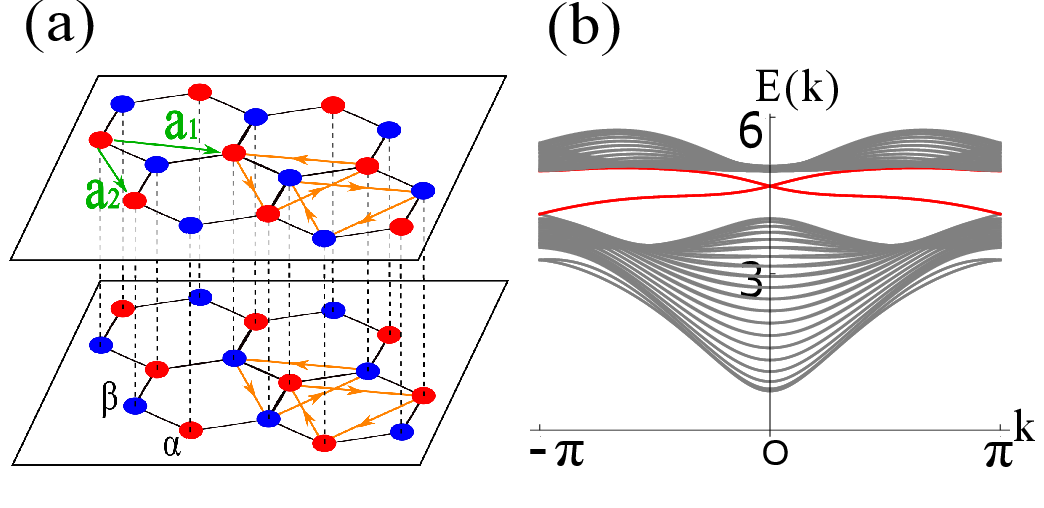}
\caption{(Color online) (a) The bilayer honeycomb system. 
%with the interactions between spins, exhibiting mSHE. 
Two sublattices are designated as $\alpha$ and $\beta$. 
The two primitive lattice vectors are represented as $\bm{a}_{1}$ and $\bm{a}_{2}$. The spins on the $\alpha$ and $\beta$ sublattices on the first (second) layer point in the $+z$ ($-z$) and $-z$ ($+z$) directions, respectively. 
The orange arrows indicate the sign convention $\xi_{ij}=+1 (=-\xi_{ji})$ for $i \to j$. 
(b) Magnon spectrum for a strip of 40 unit cells with zigzag edges 
with $J_x^{(1)} S=1.03$, $J_y^{(1)} S=0.97$, $J_z^{(1)} S=1.02$, $DS=0.2$, and $J' S=2$.
The magnon 
%helical 
edge states are shown in red.}
\label{fig:honeycomb}
\end{figure}
As a second example of MSH systems, we consider a bilayer antiferromagnetic honeycomb lattice system which does not preserve $S_z$ as in the Kane-Mele model with a finite Rashba interaction~\cite{Kane05b}.
Here, we assume the (perfect) staggered spin configuration with $N=M=1$ [see Fig.~\ref{fig:honeycomb} (a)].
The Hamiltonian of the system~\cite{comment_honeycomb1} is written as follows,
\begin{align}
&\mathcal{H}_{\rm{H}}=\mathcal{H}_{\rm{XYZ}}^{(1)}+\mathcal{H}_{\rm{XYZ}}^{(2)}+\mathcal{H}_{\rm{DM}}^{(1)}+\mathcal{H}_{\rm{DM}}^{(2)}+\mathcal{H}^{\rm (inter)},
\label{eq:honeycomb_detail} \\
&\mathcal{H}^{(l)}_{{\rm XYZ}}= \sum_{\langle i,j\rangle}   J^{(l)}_x S^{(l)}_{i,x}S^{(l)}_{j,x}  +J^{(l)}_y S^{(l)}_{i,y}S^{(l)}_{j,y}  +J^{(l)}_z S^{(l)}_{i,z}S^{(l)}_{j,z}, \nonumber \\
&\mathcal{H}_{\rm{DM}}^{(l)}=D\sum_{\langle \langle i,j \rangle \rangle} \xi_{ij} \bigl( \bm{S}_{i}^{(l)}\times\bm{S}_{j}^{(l)} \bigr)_{z},
\nonumber \\
&\mathcal{H}^{\rm (inter)}=J'\sum_{i} \bm{S}_{i}^{(1)}\cdot\bm{S}_{i}^{(2)}, \nonumber
\end{align}
where $S_{i,a}^{(l)}$ $(a=x, y, z)$ is the $a$ component of the spin at site $i$ on the $l$th layer and $J^{(1)}_{x}=J^{(2)}_y$, $J^{(1)}_y = J^{(2)}_x$, and $J^{(1)}_z=J^{(2)}_z$. 
In each layer, $\mathcal{H}_{\rm{XYZ}}^{(l)}$ describes the anisotropic (XYZ) Heisenberg interaction between nearest-neighbor spins, while $\mathcal{H}_{\rm{DM}}^{(l)}$ is the DM interaction between next-nearest-neighbor spins. 
We assume that the couplings $J_{a}^{(l)}(a=x,y,z)$ and $J'$ are all positive. 
The sign conventions of $\xi_{ij}$ are shown by the orange arrows in Fig.~\ref{fig:honeycomb}(a). 
The remaining term $\mathcal{H}^{\rm (inter)}$ represents the interlayer antiferromagnetic Heisenberg interaction. The sum runs over the vertical spin pairs shown by the dashed lines in Fig.~\ref{fig:honeycomb}(a).

Using the Holstein-Primakoff transformation,
the Hamiltonian (\ref{eq:honeycomb_detail}) is written in the same form as Eq.~(\ref{eq:H}) with
\begin{align}
&h_{1}(\bm{k})=S
\left(\begin{array}{cc}
3J_{z}^{(1)}+J'+\delta_{\bm{k}} &0  \\
0 &3J_{z}^{(1)}+J'-\delta_{\bm{k}} \\
\end{array} \right),  \nonumber \\
&h_{2}(\bm{k})=S
\left(\begin{array}{cc}
0&\frac{1}{2}(J_{x}^{(1)}-J_{y}^{(1)})\gamma_{\bm{k}}  \\
-\frac{1}{2}(J_{x}^{(1)}-J_{y}^{(1)})\gamma_{-\bm{k}} &0 \\
\end{array} \right), \nonumber \\
&\Delta_{1}(\bm{k})=S
\left(\begin{array}{cc}
J'&\frac{1}{2}(J_{x}^{(1)}+J_{y}^{(1)})\gamma_{\bm{k}}  \\
\frac{1}{2}(J_{x}^{(1)}+J_{y}^{(1)})\gamma_{-\bm{k}} &J' \\
\end{array} \right), \nonumber \\
&\Delta_{2}(\bm{k})=0, \nonumber
\end{align}
where 
%$\lambda=3J_{z}^{(1)}+J'$,
$\delta_{\bm{k}}=2D\{\sin{(\bm{k}\cdot\bm{a}_{1})}-\sin{(\bm{k}\cdot\bm{a}_{2})}-\sin{[\bm{k}\cdot(\bm{a}_{1}-\bm{a}_{2})]}\}$,
and $\gamma_{\bm{k}}=1+e^{{\rm i}\,\bm{k}\cdot\bm{a}_{1}}+e^{{\rm i}\,\bm{k}\cdot\bm{a}_{2}}$. 
Note that $h_{2}(\bm{k})$ does not conserve $S_z$. 
%violates the conservation of $S_{z}$. 
As shown in Fig. \ref{fig:honeycomb}(a), $\bm{a}_{1}$ and $\bm{a}_{2}$ are the primitive lattice vectors of the honeycomb lattice. 

Figure~\ref{fig:honeycomb}(b) shows the magnon dispersion in a strip of the bilayer honeycomb lattice with zigzag edges. 
The helical edge states (shown in red) exist and traverse the energy gap, as in the kagome bilayer system.
Applying Eq.~(\ref{eq:D}) to the system, we find that the ${\mathbb Z}_2$ index of each magnon band is unity, namely, $D_{n}=1$ for $n=1,2$~\cite{comment_BZ,comment_honeycomb2}, reflecting the existence of the helical edge states. These helical edge states are expected to be responsible for 
%related to 
the magnon spin Nernst effect studied in Ref.~\cite{Zyuzin16} if the XYZ term $\mathcal{H}^{(l)}_{\rm XYZ}$ is almost isotropic. 
The topological invariants are unchanged for any parameters as long as the staggered order is stable. 
One can construct an example of a topologically trivial phase with $D_{n} = 0$ for $n=1,2$ by adding a single-ion anisotropy term to the Hamiltonian Eq. (\ref{eq:honeycomb_detail}). See Supplemental Material for details.

%Summary
{\it Summary.}
In summary, we have defined the ${\mathbb Z}_2$ index for the
magnon spin Hall systems. We have also demonstrated the validity and robustness of the index in two cases, ``ferromagnetic'' kagome and antiferromagnetic honeycomb bilayer systems.
We found that the value of the invariant $D_{n}=1$ $(0)$ 
characterizes the presence (absence) of the nontrivial edge states. 
It is worth noting that the expression of the ${\mathbb Z}_2$ index is applicable even in the system without the conservation of $S_{z}$, i.e., the latter case. 
%This ${\mathbb Z}_2$ %topological number 
%index can be applied not only to magnons, but also to other bosonic systems. %such as photons and phonons. 
%This ${\mathbb Z}_2$ topological number characterizes the topological phase of the magnon spin Hall insulator. 
%A pair of states with the opposite %magnetic dipole moment 
%angular momenta propagating in the opposite directions along the edges is characterized as $D_{n\sigma}=1$. %, while $D_{n\sigma}=0$ for a pair of topologically trivial states.
In addition, the topological phases of magnons in the proposed bilayer systems are robust against disorder in the interlayer antiferromagnetic couplings, as it does not break the pseudo-time-reversal symmetry $\Theta'$. 

The proposed ${\mathbb Z}_2$ index will be useful for identifying the magnetic counterpart of class AII topological insulators in a wide variety of materials. 
Since various methods for measuring magnon current have been proposed~\cite{Shiomi17}, 
%such as the one using the inverse spin Hall effect by converting magnon current into electrical spin current~\cite{Shiomi17}, 
we expect the helical magnon edge states to be observed in real materials in the near future using currently available experimental techniques. Finally, it should be noted that the ${\mathbb Z}_2$ index can be applied not only to magnons, but also to other bosonic quasiparticles. 
Therefore, our work can pave the way for further studies on other classes of bosonic topological phases.

{\it Acknowledgements.}
This work was supported by JSPS KAKENHI Grants No. JP17K14352, No. JP18H04478, No. JP18K03445, and No. JP18H04220.
H.K. was supported by the  JSPS through Program for Leading Graduate Schools (ALPS).

\begin{widetext}
%Supplementary Materials
\begin{center}
\Large{Supplemental Material for: {\it ${\mathbb Z}_2$ Topological Invariant for Magnon Spin Hall Insulator} }
\end{center}

\begin{center}
\bf{The existence of a Kramers pairs of bosons}
\end{center}
In the main text, we introduce the pseudo-time-reversal operator $\Theta'=PK$ which satisfies Eqs.~(\ref{eq:Theta})-(\ref{eq:P}).
In this section, we show that %this
the operator ensures the existence of ``Kramers pairs" of bosons.
%First, let us multiply the both sides of the eigen-equation of the BdG Hamiltonian:
%To begin with, let us multiply $\Theta'(\bm{k})$ 
First, let us begin by the following eigen-equation of the BdG Hamiltonian:
\begin{align}
\Sigma_{z}H(\bm{k})\bm{\psi}(\bm{k})=E(\bm{k})\bm{\psi}(\bm{k}).
\label{eq:BdG_sup}
\end{align}
Multiplying $\Theta'$ from the left side of Eq.~(\ref{eq:BdG_sup}), we obtain
%by the matrices $\Theta'(\bm{k})$. Using Eq.~(\ref{eq:C}), we obtain
\begin{align}
\Sigma_{z}H(-\bm{k})\Theta'\bm{\psi}(\bm{k})=E(\bm{k})\Theta'\bm{\psi}(\bm{k}).
\label{eq:BdG_TR}
\end{align}
Here we used Eq.~(\ref{eq:Commutation}) in the main text. 
At the time-reversal-invariant momenta (TRIM) 
$\bm{k}=\bm{\Lambda}$, the two vectors $\bm{\psi}(\bm{\Lambda})$ and $\Theta'(\bm{\Lambda})\bm{\psi}(\bm{\Lambda})$ are eigenvectors of $\Sigma_{z}H(\bm{\Lambda})$ with the same eigenvalue $E(\bm{\Lambda})$, which follows from Eq.~(\ref{eq:BdG_TR}). 
In the following, we show that these two vectors are %paraorthogonal. 
orthogonal to each other at TRIM. The inner product 
of $\bm{\phi}(-\bm{k})$ and $\Theta'\bm{\psi}(\bm{k})$
%can be deformed as
yields
\begin{align}
%\left\langle \! \! \left\langle \bm{\phi}(-\bm{k}),%\Theta'(\bm{k})\bm{\psi}(\bm{k})
%\bm{\psi}'(\bm{k}) \right\rangle \! \! \right\rangle 
\left\langle \!  \left\langle \bm{\phi}(-\bm{k}),\Theta'\bm{\psi}(\bm{k})\right\rangle \! \right\rangle 
&=\phi_{i}^{*}(-\bm{k})\left(\Sigma_{z}P\right)_{ij}\psi_{j}^{*}(\bm{k}) \nonumber \\
&=\psi_{j}^{*}(\bm{k})\left(\Sigma_{z}P\right)_{ji}^{T}\phi_{i}^{*}(-\bm{k}) \nonumber \\
&=\left\langle \bm{\psi}(\bm{k}), P^{T}\Sigma_{z}K\bm{\phi}(-\bm{k}) \right\rangle \nonumber \\
&=\left\langle \! \left\langle \bm{\psi}(\bm{k}), \Sigma_{z}P^{T}\Sigma_{z}K\bm{\phi}(-\bm{k})\right\rangle \! \right\rangle.
\label{eq:Kramers1}
\end{align}
In the third line of Eq.~(\ref{eq:Kramers1}), the inner product is defined as $\langle \bm{\phi},\bm{\psi} \rangle=\bm{\phi}^{\dagger}\bm{\psi}$.
%which is the same as that in fermionic systems.
%Here, the inner product is defined as $\langle \bm{\phi},\bm{\psi} \rangle=\bm{\phi}^{\dagger}\bm{\psi}$, which is the same as that in fermionic systems.
By replacing $\bm{\phi}(-\bm{k})$ with $\Theta'\bm{\phi}(-\bm{k})$, %one finds
the inner product can be rewritten into the following form:
\begin{align}
\left\langle \! \left\langle \Theta'\bm{\phi}(-\bm{k}),\Theta'\bm{\psi}(\bm{k})\right\rangle \! \right\rangle 
&=\left\langle \! \left\langle \bm{\psi}(\bm{k}), \Sigma_{z}P^{T}\Sigma_{z}KPK\bm{\phi}(-\bm{k})\right\rangle \! \right\rangle \nonumber \\
&=\left\langle \! \left\langle \bm{\psi}(\bm{k}), \bm{\phi}(-\bm{k})\right\rangle \! \right\rangle,
\label{eq:Kramers2}
\end{align}
where we used Eq.~(\ref{eq:P}) in the main text. Then one finds that the inner product of $\bm{\psi}(\bm{k})$ and $\Theta'\bm{\psi}(-\bm{k})$ satisfies
\begin{align}
\left\langle \! \left\langle \bm{\psi}(\bm{k}),\Theta'\bm{\psi}(-\bm{k})\right\rangle \!\right\rangle 
&=\left\langle \! \left\langle \Theta'^{2}\bm{\psi}(-\bm{k}), \Theta'\bm{\psi}(\bm{k})\right\rangle \! \right\rangle \nonumber \\
&=-\left\langle \! \left\langle \bm{\psi}(-\bm{k}),\Theta'\bm{\psi}(\bm{k})\right\rangle \! \right\rangle.
\label{eq:Kramers3}
\end{align}
%Here, we used Eq.~(\ref{eq:Kramers2}) by substituting $\bm{\phi}(-\bm{k})$ for $\Theta'\bm{\psi}(-\bm{k})$.
It follows from Eq.~(\ref{eq:Kramers3}) that the two vectors $\bm{\psi}(\bm{\Lambda})$ and $\Theta'\bm{\psi}(\bm{\Lambda})$ at the TRIM ($\bm{k}=\bm{\Lambda}$) are orthogonal, 
%we find that the two vectors $\bm{\psi}(\bm{\Lambda})$ and $\Theta'\bm{\psi}(\bm{\Lambda})$ with the same energy $E(\bm{\Lambda})$ are orthogonal, 
i.e.,
\begin{align}
\left\langle \! \left\langle \bm{\psi}(\bm{\Lambda}), \Theta'\bm{\psi}(\bm{\Lambda})\right\rangle \! \right\rangle=0.
\label{eq:Kramers4}
\end{align}
%Then
%Finally, we conclude that 
Therefore, the ``Kramers pairs'' of bosons %exists 
can be defined under the condition~(\ref{eq:Theta})-(\ref{eq:Commutation}) in the main text. %that
%where the operator $\Theta'(\bm{k})$ satisfies Eqs.~(\ref{eq:C})-(\ref{eq:P}).

\begin{center}
\bf{Berry curvature and Berry connection of bilayer ``ferromagnet"}
\end{center}
In the bilayer of ``ferromagnetic'' general lattice systems without net-moment (one of the examples is the bilayer kagome system described by Fig.~\ref{fig:kagome} in the main text),
we show that the Berry connections and curvatures in the bilayer perfectly coincide with those of the two independent single layer systems without the interlayer coupling $J'$.  
The BdG Hamiltonian of the ``ferromagnetic'' bilayer system %always 
takes the same form as Eq.~(\ref{eq:H}), i.e., %with $G(\bm{k})=H^{(single)}(\bm{k})+J'S1_{N}$ and $F(\bm{k})=J'S1_{N}$,
\begin{align}
H(\bm{k})=
\left(
\begin{array}{cccc}
H^{(\rm{single})}(\bm{k})+J'S1_{N} &0 &0 &J'S1_{N} \\
0 &H^{(\rm{single})*}(-\bm{k})+J'S1_{N}  &J'S1_{N}&0 \\
0 &J'S1_{N} &H^{(\rm{single})*}(-\bm{k})+J'S1_{N} &0 \\
J'S1_{N} &0 &0 &H^{(\rm{single})}(\bm{k})+J'S1_{N} \\
\end{array} 
\right),
\end{align}
where $H^{(\rm{single})}(\bm{k})$ is the Hamiltonian of the ferromagnetic single layer system.
To diagonalize the Hamiltonian with the paraunitary matrix $T(\bm{k})$ which satisfies $T^{\dagger}(\bm{k})\Sigma_{z}T(\bm{k})=\Sigma_{z}$, we need to solve the eigenvalue problem of $\Sigma_{z}H(\bm{k})$:
\begin{align}
\Sigma_{z}H(\bm{k})\bm{\Psi}_{n,l,\sigma}(\bm{k})=E_{n,l,\sigma}(\bm{k})\bm{\Psi}_{n,l,\sigma}(\bm{k}).
\label{eq:BdG_GF}
\end{align}
Using the eigenvector $\bm{\psi}_{n}(\bm{k})$ and eigenvalue $\lambda_{n}(\bm{k})$ of the single layer Hamiltonian $H^{(\rm{single})}(\bm{k})$, the eigenvalues $E_{n,l,\sigma}(\bm{k})$ 
and the eigenvectors $\bm{\Psi}_{n,l,\sigma}(\bm{k})$ in Eq.~(\ref{eq:BdG_GF}) can be written as
\begin{align}
&E_{n,1,\sigma}(\bm{k})=\sigma\sqrt{\left({\lambda}_{n}(\sigma\bm{k})+J'S\right)^2-(J'S)^2},  \\
&E_{n,2,\sigma}(\bm{k})=\sigma\sqrt{\left({\lambda}_{n}(-\sigma\bm{k})+J'S\right)^2-(J'S)^2}, \\
&\bm{\Psi}_{n,1,+}(\bm{k})=
\left(
\begin{array}{cccc}
\cosh{(\theta_{n}(\bm{k}))}\bm{\psi}_{n}(\bm{k}) \\
0  \\
0  \\
\sinh{(\theta_{n}(\bm{k}))}\bm{\psi}_{n}(\bm{k}) \\
\end{array} 
\right),  \label{eq:1+}\\
&\bm{\Psi}_{n,2,+}(\bm{k})=
\left(
\begin{array}{cccc}
0 \\
\cosh{(\theta_{n}(-\bm{k}))}\bm{\psi}_{n}^{*}(-\bm{k})  \\
\sinh{(\theta_{n}(-\bm{k}))}\bm{\psi}_{n}^{*}(-\bm{k})  \\
0 
\end{array} 
\right),  \label{eq:2+}\\
&\bm{\Psi}_{n,1,-}(\bm{k})=
\left(
\begin{array}{cccc}
0 \\
\sinh{(\theta_{n}(-\bm{k}))}\bm{\psi}_{n}^{*}(-\bm{k})  \\
\cosh{(\theta_{n}(-\bm{k}))}\bm{\psi}_{n}^{*}(-\bm{k})  \\
0 
\end{array} 
\right),  \label{eq:1-}\\
&\bm{\Psi}_{n,2,-}(\bm{k})=
\left(
\begin{array}{cccc}
\sinh{(\theta_{n}(\bm{k}))}\bm{\psi}_{n}(\bm{k}) \\
0  \\
0  \\
\cosh{(\theta_{n}(\bm{k}))}\bm{\psi}_{n}(\bm{k}) 
\end{array} 
\right).\label{eq:2-}
\end{align}
Here, $\theta_{n}(\bm{k})$ is defined as
\begin{align}
\tanh{\left(\theta_{n}(\bm{k})\right)}=\frac{-\left(\lambda_{n}(\bm{k})+J'S\right)+\sqrt{\left(\lambda_{n}(\bm{k})+J'S\right)^2-(J'S)^2}}{J'S}.
\end{align}
Then the paraunitary matrix $T(\bm{k})$ and the diagonalized Hamiltonian are %written as
given by
\begin{align}
&T(\bm{k})
=\left(
\bm{\Psi}_{1,1,+}(\bm{k}),\cdots,\bm{\Psi}_{N,1,+}(\bm{k}),
\bm{\Psi}_{1,2,+}(\bm{k}),\cdots,\bm{\Psi}_{N,2,+}(\bm{k}), 
\bm{\Psi}_{1,1,-}(\bm{k}),\cdots,\bm{\Psi}_{N,1,-}(\bm{k}),
\bm{\Psi}_{1,2,-}(\bm{k}),\cdots,\bm{\Psi}_{N,2,-}(\bm{k})
\right), \\
&T^{\dagger}(\bm{k})H(\bm{k})T(\bm{k}) \nonumber \\ 
&={\rm diag}\left[
E_{1,1,+}(\bm{k}),\cdots \!,E_{N,1,+}(\bm{k}),E_{1,2,+}(\bm{k}),\cdots \!,E_{N,2,+}(\bm{k}),
-E_{1,1,-}(\bm{k}),\cdots \!,-E_{N,1,-}(\bm{k}),-E_{1,2,-}(\bm{k}),\cdots \!,-E_{N,2,-}(\bm{k})
\right]. 
\end{align}
In the limit of $J'\rightarrow 0$, $\bm{\Psi}_{n,l,\sigma}(\bm{k})$ coincides with the $n$th eigenstate of the $l$th layer. 
The two eigenvectors $\bm{\Psi}_{n,1,\sigma}(\bm{k})$ and $\bm{\Psi}_{n,2,\sigma}(\bm{k})$ form a ``Kramers pair'' of magnons. 
The Berry connection and curvature of bosonic system described by the BdG Hamiltonian~\cite{Matsumoto14} is defined by
\begin{align}
&\bm{A}_{n,l,\sigma}(\bm{k})
:= {\rm i}\, \sigma \left\langle \! \left\langle  \bm{\Psi}_{n,l,\sigma}(\bm{k}),\nabla_{\bm{k}}\bm{\Psi}_{n,l,\sigma}(\bm{k})  \right\rangle \! \right\rangle, \label{eq:A}\\
&\Omega_{n,l,\sigma}(\bm{k}):= \bigl( \nabla_{\bm{k}} \times \bm{A}_{n,l,\sigma}(\bm{k}) \bigr)_z.\label{eq:Omega}
\end{align} 
By substituting Eqs.~(\ref{eq:1+})-(\ref{eq:2-}) into Eq.~(\ref{eq:A}), %one finds
we find the following relations:
\begin{align}
&\bm{A}_{n,1,+}(\bm{k}) =\bm{A}_{n,2,-}(\bm{k})\nonumber \\
&={\rm i}\,\left\langle\cosh{(\theta_{n}(\bm{k}))}\bm{\psi}_{n}(\bm{k}),
\nabla_{\bm{k}}\cosh{(\theta_{n}(\bm{k}))}\bm{\psi}_{n}(\bm{k})\right\rangle
- {\rm i}\,\left\langle\sinh{(\theta_{n}(\bm{k}))}\bm{\psi}_{n}(\bm{k}),
\nabla_{\bm{k}}\sinh{(\theta_{n}(\bm{k}))}\bm{\psi}_{n}(\bm{k})\right\rangle \nonumber \\
&={\rm i}\,\left\langle\bm{\psi}_{n}(\bm{k}),\nabla_{\bm{k}}\bm{\psi}_{n}(\bm{k})\right\rangle\nonumber \\
&=\bm{A}_{n}^{(\rm{single})}(\bm{k}),
\\
&\bm{A}_{n,2,+}(\bm{k}) =\bm{A}_{n,1,-}(\bm{k})\nonumber \\
&={\rm i}\,\left\langle\cosh{(\theta_{n}(-\bm{k}))}\bm{\psi}_{n}(-\bm{k}),
\nabla_{\bm{k}}\cosh{(\theta_{n}(-\bm{k}))}\bm{\psi}_{n}(-\bm{k})\right\rangle^{*}
-{\rm i}\,\left\langle\sinh{(\theta_{n}(-\bm{k}))}\bm{\psi}_{n}(-\bm{k}),
\nabla_{\bm{k}}\sinh{(\theta_{n}(-\bm{k}))}\bm{\psi}_{n}(-\bm{k})\right\rangle^{*}\nonumber \\
&={\rm i}\,\left\langle\bm{\psi}_{n}(-\bm{k}),\nabla_{\bm{k}}\bm{\psi}_{n}(-\bm{k})\right\rangle^{*} \nonumber \\
&={\rm i}\,\left\langle\bm{\psi}_{n}(-\bm{k}),\nabla_{-\bm{k}}\bm{\psi}_{n}(-\bm{k})\right\rangle\nonumber \\
&=\bm{A}_{n}^{(\rm{single})}(-\bm{k}),
\end{align}
where $\bm{A}_{n}^{(\rm{single})}(\bm{k})={\rm i}\,\left\langle\bm{\psi}_{n}(\bm{k}),\nabla_{\bm{k}}\bm{\psi}_{n}(\bm{k})\right\rangle$ is the Berry connection of the single layer system.
Using the Berry curvature of the single layer system %$\Omega_{n}^{(single)}(\bm{k})=\nabla_{\bm{k}} \times \bm{A}_{n}^{(single)}(\bm{k})$
 $\Omega_{n}^{({\rm single})}(\bm{k})=\bigl( \nabla_{\bm{k}} \times \bm{A}_{n}^{(\rm{single})}(\bm{k}) \bigr)_z$, the Berry curvature (\ref{eq:Omega}) can be written as
\begin{align}
&\Omega_{n,1,+}(\bm{k})=\Omega_{n,2,-}(\bm{k})=\Omega_{n}^{(\rm{single})}(\bm{k}), \nonumber \\
&\Omega_{n,2,+}(\bm{k})=\Omega_{n,1,-}(\bm{k})=-\Omega_{n}^{(\rm{single})}(-\bm{k}). \label{eq:Berry_bilayer_singlelayer}
\end{align}
We conclude that the Berry connection and curvature of the ``ferromagnetic'' bilayer systems do not depend on the interlayer coupling $J'$ and are the same as %that
those of the two independent single layer systems. 
The Chern number of each band always takes zero due to $\Omega_{n,1,\sigma}(\bm{k}) + \Omega_{n,2,\sigma}(-\bm{k})=0$ from Eq.~(\ref{eq:Berry_bilayer_singlelayer}), leading to no thermal Hall effect.
We emphasize that Eq.~(\ref{eq:Berry_bilayer_singlelayer}) is valid even in general ``ferromagnetic" bilayer systems where spins point in the same direction on the same layer while spins on the two layers face each other.
Therefore, we can simply construct the systems with MSH effect via making bilayer from two single layers, each of which exhibits the thermal Hall effect.

\begin{center}
\bf{Demonstration in topologically trivial phases}
\end{center}
In this section, we show the results of bilayers of kagome and honeycomb systems with trivial phase and confirm that our ${\mathbb Z}_2$ index can characterize the phase.
First, we consider the bilayer kagome lattice system without the DM interaction and assume the same magnetic pattern as in Fig.~\ref{fig:kagome}(a). The Hamiltonian of this system is written as
\begin{align}
\mathcal{H}_{\rm{K}}=\mathcal{H}^{(1)}+\mathcal{H}^{(2)}+\mathcal{H}^{\rm (inter)}. 
\end{align}
Here, $\mathcal{H}^{(l)}$ and $\mathcal{H}^{\rm (inter)}$ are defined by
\begin{align}
&\mathcal{H}^{(l)}=-J\sum_{\langle i,j\rangle} \bm{S}_{i}^{(l)}\cdot\bm{S}_{j}^{(l)}-\kappa_{1}\sum_{i\in\beta}(S_{i,z}^{(l)})^{2}-\kappa_{2}\sum_{i\in\gamma}(S_{i,z}^{(l)})^{2},
\nonumber \\
&\mathcal{H}^{\rm (inter)}=J'\sum_{i} \bm{S}_{i}^{(1)}\cdot\bm{S}_{i}^{(2)}.\label{eq:kagome_trivial}
\end{align}
The second and third terms of $\mathcal{H}^{(l)}$ are the easy axis anisotropy on the sublattice $\beta$ and $\gamma$, respectively.
Applying the Holstein-Primakoff transformation to spins $\bm{S}_{i}^{(l)}$,
the Hamiltonian (\ref{eq:kagome_trivial}) takes the same form as Eq.~(\ref{eq:H}) with
\begin{align}
h_{1}(\bm{k})=
S\left(
\begin{array}{ccc}
4J+J' &-J p_{1}^{*} &-J p_{2}^{*} \\
-J p_{1} &4J+J'+2\kappa_{1} &-J p_{3}^{*}  \\
-J p_{2} &-J p_{3} &4J+J'+2\kappa_{2}  \\
\end{array} 
\right),
\end{align}
$h_{2}(\bm{k})=0,\Delta_{1}(\bm{k})=J'S1_{3}$, and $\Delta_{2}(\bm{k})=0$, where $p_i=1+\exp({\rm i}\,{\bm k}\cdot {\bm a}_i)$ ($i=1,2,3$) with ${\bm a}_3:={\bm a}_2-{\bm a}_1$.

The second case is the bilayer honeycomb lattice system without the DM interaction. We assume that the magnetic ordering is the same as that in Fig.~\ref{fig:honeycomb}(a).
The Hamiltonian of this system is written as
\begin{align}
&\mathcal{H}_{\rm{H}}=\mathcal{H}^{(1)}+\mathcal{H}^{(2)}+\mathcal{H}^{\rm (inter)},
\nonumber \\
&\mathcal{H}^{(l)}=\sum_{\langle i,j\rangle} J_{x}^{(l)}S_{i,x}^{(l)}S_{j,x}^{(l)}+J_{y}^{(l)}S_{i,y}^{(l)}S_{j,y}^{(l)}+J_{z}^{(l)}S_{i,z}^{(l)}S_{j,z}^{(l)}-\kappa\sum_{i\in\beta}(S_{i,z}^{(l)})^{2},
\nonumber \\
&\mathcal{H}^{\rm (inter)}=J'\sum_{i} \bm{S}_{i}^{(1)}\cdot\bm{S}_{i}^{(2)}.
\label{eq:honeycomb_detail_trivial}
\end{align}
Using the Holstein-Primakoff transformation,
the Hamiltonian (\ref{eq:honeycomb_detail_trivial}) is written as the same form as Eq.~(\ref{eq:H}) with
\begin{align}
&h_{1}(\bm{k})=S
\left(\begin{array}{cc}
3J^{(1)}_z+J' &0  \\
0 & 3J^{(1)}_z+J' +2\kappa \\
\end{array} \right),  \nonumber \\
&h_{2}(\bm{k})=S
\left(\begin{array}{cc}
0&(J_{x}^{(1)}-J_{y}^{(1)})\gamma_{\bm{k}}  \\
-(J_{x}^{(1)}-J_{y}^{(1)})\gamma_{-\bm{k}} &0 \\
\end{array} \right), \nonumber \\
&\Delta_{1}(\bm{k})=S
\left(\begin{array}{cc}
J'&\frac{1}{2}(J_{x}^{(1)}+J_{y}^{(1)})\gamma_{\bm{k}}  \\
\frac{1}{2}(J_{x}^{(1)}+J_{y}^{(1)})\gamma_{-\bm{k}} &J' \\
\end{array} \right), \nonumber \\
&\Delta_{2}(\bm{k})=0,
\end{align}
where $J_{x}^{(1)}=J_{y}^{(2)},J_{y}^{(1)}=J_{x}^{(2)}$, $J_{z}^{(1)}=J_{z}^{(2)}$, and $\gamma_{\bm{k}}=1+e^{{\rm i}\,\bm{k}\cdot\bm{a}_{1}}+e^{{\rm i}\,\bm{k}\cdot\bm{a}_{2}}$. 
As in the previous case, we removed the DM interaction and added the easy axis anisotropy term on $\beta$ sublattice.
The energy spectra of these systems are shown in Fig.~\ref{fig:trivial}. There are no topologically protected edge states in either case.
\begin{figure}[H]
  \centering
  \includegraphics[width=12.5cm]{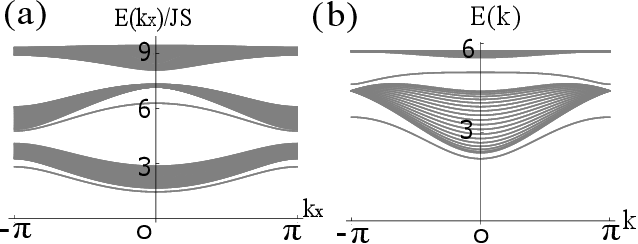}
\caption{(a) Magnon spectrum of the strip of the bilayer kagome lattice expressed in Eq.~(\ref{eq:kagome_trivial}) with $M=20$ width for $\kappa_{1}=J$, $\kappa_{2}=2J$, $J'=0.1J$. 
(b) Magnon spectrum of the strip of the bilayer honeycomb lattice represented as Eq.~(\ref{eq:honeycomb_detail_trivial}) of 40 unit cells width with the zigzag edges. The parameters are chosen to be $J_x^{(1)} S=1.03$, $J_y^{(1)} S=0.97$, $J_z^{(1)} S=1.02$, $\kappa S=0.5$ and $J' S=2$.
In both cases, 
there are the edge states which do not traverse the energy gaps, 
i.e., topologically unprotected edge states.}
\label{fig:trivial}
\end{figure}
\noindent
Correspondingly, using Eq.~(\ref{eq:D}), we obtain that all the $\mathbb{Z}_{2}$ indices are zero in both systems.

\end{widetext}


\begin{thebibliography}{99}
\bibitem{Chern_hole} 
The Chern numbers of the three hole bands %in the hole space 
are opposite to those of the corresponding particle bands.
\bibitem{comment_honeycomb1} The model can be regarded as that proposed by Zyuzin and Kovalev~\cite{Zyuzin16} with a modification, i.e., the Heisenberg interaction is replaced with
an anisotropic Heisenberg interaction.
\bibitem{comment_BZ}
As in the case of the kagome-lattice system, we consider the deformed honeycomb lattice whose Brillouin zone is a square.
\bibitem{comment_honeycomb2} 
Unlike the first example, 
%case of the kagome ferromagnet, 
the Berry connections and curvatures of the bilayer antiferromagnetic honeycomb lattice cannot be reduced to those of the single-layer system.
\end{thebibliography}
\end{document}